\documentclass[notitlepage,floatfix,aps,prl,reprint,amsmath,amssymb,twocolumn,superscriptaddress,10pt]{revtex4-1}
\usepackage{graphicx}
\usepackage{dcolumn}
\usepackage{bm}
\usepackage{braket}
\usepackage{amssymb}
\usepackage{hyperref}
\usepackage{amsmath}
\usepackage{textcomp}
\usepackage{amsfonts}
\usepackage{gensymb}

\begin{document}

\title{Coherent Dynamics in Quantum Emitters under Dichromatic Excitation}
\author{Z. X. Koong}
\email[Correspondence: ]{zk49@hw.ac.uk}
\affiliation{
 SUPA, Institute of Photonics and Quantum Sciences, Heriot-Watt University, EH14 4AS, United Kingdom
}
\author{E. Scerri}
\email[Current address: Leiden University, P.O. Box 9504, 2300 RA Leiden, The Netherlands]{}
\affiliation{
 SUPA, Institute of Photonics and Quantum Sciences, Heriot-Watt University, EH14 4AS, United Kingdom
}
\author{M. Rambach}
\email[Current address: Centre for Engineered Quantum Systems, School of Mathematics and Physics, University of Queensland, QLD 4072, Australia]{}
\affiliation{
 SUPA, Institute of Photonics and Quantum Sciences, Heriot-Watt University, EH14 4AS, United Kingdom
}
\author{M. Cygorek}
\affiliation{
 SUPA, Institute of Photonics and Quantum Sciences, Heriot-Watt University, EH14 4AS, United Kingdom
}
\author{M. Brotons-Gisbert}
\affiliation{
 SUPA, Institute of Photonics and Quantum Sciences, Heriot-Watt University, EH14 4AS, United Kingdom
}
\author{R. Picard}
\affiliation{
 SUPA, Institute of Photonics and Quantum Sciences, Heriot-Watt University, EH14 4AS, United Kingdom
}
\author{Y. Ma}
\affiliation{
College of Optoelectronic Engineering, Chongqing University of Posts and Telecommunications, Chongqing 400065, China
}
\author{S. I. Park}
\affiliation{
 Center for Opto-Electronic Materials and Devices Research, Korea Institute of Science and Technology, Seoul 02792, Republic of Korea
}
\author{J. D. Song}
\affiliation{
 Center for Opto-Electronic Materials and Devices Research, Korea Institute of Science and Technology, Seoul 02792, Republic of Korea
}
\author{E. M. Gauger}
\author{B. D. Gerardot}
\email[Correspondence: ]{b.d.gerardot@hw.ac.uk}
\affiliation{
 SUPA, Institute of Photonics and Quantum Sciences, Heriot-Watt University, EH14 4AS, United Kingdom
}

\date{\today}
\begin{abstract}
We characterize the coherent dynamics of a two-level quantum emitter driven by a pair of symmetrically-detuned phase-locked pulses.
The promise of dichromatic excitation is to spectrally isolate the excitation laser from the quantum emission, enabling background-free photon extraction from the emitter.
Paradoxically, we find that excitation is not possible without spectral overlap between the exciting pulse and the quantum emitter transition for ideal two-level systems due to cancellation of the accumulated pulse area.  
However, any additional interactions that interfere with cancellation of the accumulated pulse area may lead to a finite stationary population inversion.
Our spectroscopic results of a solid-state two-level system show that while coupling to lattice vibrations helps to improve the
inversion efficiency up to 50\% under symmetric driving, coherent population control and a larger amount of inversion are possible using asymmetric dichromatic excitation, which we achieve by adjusting the ratio of the intensities between the red and blue-detuned pulses.
Our measured results, supported by simulations using a real-time path-integral method, offer a new perspective towards realizing efficient, background-free photon generation and extraction.
\end{abstract}

\maketitle

Solid-state quantum emitters, in particular semiconductor quantum dots (QD), offer a promising platform for generating quantum states that can facilitate dephasing-free information transfer between nodes within an optical quantum network~\cite{obrien_photonic_2009,de_riedmatten_long_2004,hensen_loophole-free_2015,liao_satellite--ground_2017}. 
On-demand indistinguishable photon streams for this purpose can be made using coherent excitation of QDs~\cite{senellart_high-performance_2017}. 
While resonance fluorescence of QD suppresses detrimental environmental charge noise~\cite{kuhlmann_transform-limited_2015} and timing jitter~\cite{kambs_limitations_2018} in the photon emission, the excitation laser must be filtered from the single photon stream. 
Typically, this is achieved with polarization filtering of the resonant laser.
However, unless employing a special microcavity design~\cite{gerhardt_intrinsic_2018,wang_towards_2019-1,tomm_bright_2020}, polarization filtering inherently reduces collection efficiency by at least 50\%. 
This motivates the consideration of alternative off-resonant excitation techniques to allow spectral filtering~\cite{gustin_efficient_2019}.
In particular, off-resonant phonon-assisted excitation~\cite{reindl_highly_2019,thomas_efficient_2020}, resonant two-photon excitation~\cite{liu_solid-state_2019} and resonant Raman excitation~\cite{he_indistinguishable_2013} schemes benefit from being able to spectrally isolate the zero-phonon line from the laser spectrum, to enable efficient single photon generation.

\begin{figure}
    \includegraphics[width=0.5\textwidth]{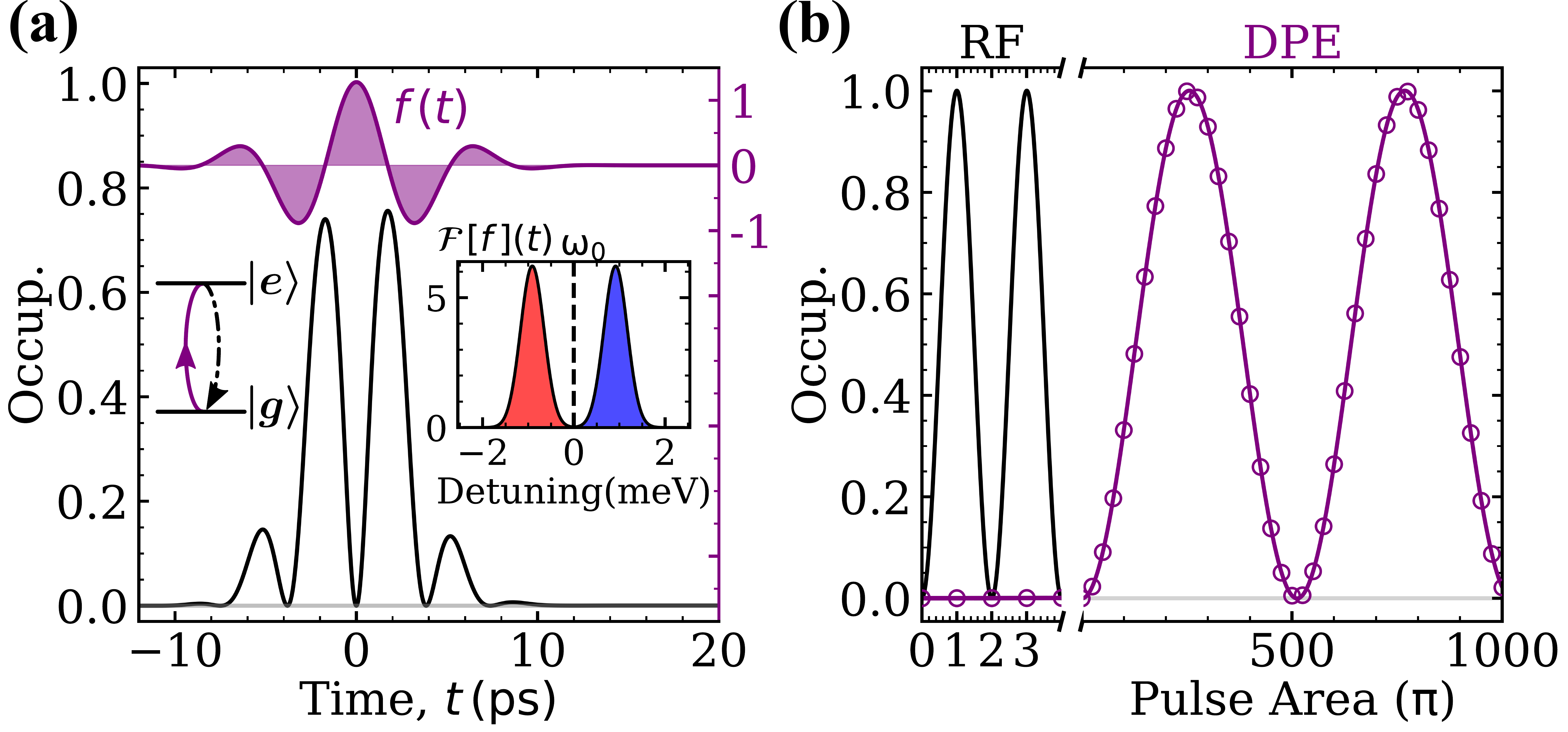}
    \caption{\textbf{Symmetric dichromatic excitation of a two-level system.}
    \textbf{(a)} Temporal evolution of the symmetric dichromatic excitation field, $f(t)$ (purple) and the associated spectrum (inset, $\mathcal{F}[f](t)$).
    The corresponding excited state occupation (black) upon interaction with the dissipation-free 2LS highlights the vanishing population inversion as a result of the net cancellation of the temporal pulse area.
    \textbf{(b)} The excited state occupation as a function of pulse area of the driving field for both monochromatic resonant driving (resonance fluorescence, RF) and dichromatic pulse excitation (DPE). 
    Results from numerical simulation (purple circles) match well with our analytical expression (solid line).
    }
    \label{fig:1s}
\end{figure}

However, the benefits of these non-resonant schemes are accompanied by their intrinsic drawbacks (e.g.~emission time jitter~\cite{simon_creating_2005,huber_measurement_2013}, nuclear spin noise~\cite{he_indistinguishable_2013,malein_screening_2016,sun_measurement_2016} and excitation-induced dephasing~\cite{ramsay_damping_2010,monniello_excitation_2013}) which inevitably degrade the photon indistinguishability.
To address these issues, He \textit{et al.}~\cite{he_coherently_2019-1} propose a coherent driving scheme using a pair of pulses, each with envelope $\epsilon(t)$ and detuned by $\pm \Delta$ from the fundamental transition of the emitter $\omega_0$.
The idea behind such dichromatic excitation is that the combined effect $f(t)$ of two identical, equally detuned pulses with envelopes $\epsilon(t)$ becomes equivalent to a single resonant pulse with modified envelope $\epsilon'(t)=2\epsilon(t)\cos(\Delta t)$:
\begin{align}
f(t)&=\epsilon(t)\cos\big((\omega_0-\Delta) t)\big) +
\epsilon(t)\cos\big((\omega_0+\Delta) t\big) \nonumber \\
&= 2\epsilon(t)\cos(\Delta t) \cos(\omega_0 t) = \epsilon'(t)  \cos(\omega_0 t)  ~.
\label{eq:addition}
\end{align}
The argument is that this would make it possible to efficiently excite quantum emitters using pulses that are spectrally separated from the fundamental transition.

However, a different picture unfolds when the system dynamics is considered in more detail: the Hamiltonian of an ideal two-level system (2LS) driven by a dichromatic pulse $f(t)=\epsilon_R(t)e^{-i\Delta t}+\epsilon_B(t)e^{i\Delta t}$
with real envelopes of the red- and blue-detuned pulses $\epsilon_R(t)$ and
$\epsilon_B(t)$ (in the rotating frame with respect to the 2LS) is given by
\begin{align}
    H=&\frac{\hbar}2 \big(f(t)^* |e\rangle\langle g| + f(t) |g\rangle\langle e|\big)
    = \boldsymbol{\Omega}(t)\cdot \mathbf{s} ~,
\\
\boldsymbol{\Omega}(t)=&\left(\begin{array}{c}
\textrm{Re}\{f(t)\} \\ \textrm{Im}\{f(t)\} \\ 0
\end{array}\right)=
\left(\begin{array}{c}
\big(\epsilon_R(t)+\epsilon_B(t)\big) \cos(\Delta t)\\
-\big(\epsilon_R(t)-\epsilon_B(t)\big) \sin(\Delta t)  \\ 0
\end{array}\right) ~,
\end{align}  
where we have expressed the two-level state as a pseudo-spin Bloch vector $\mathbf{s}$ precessing about the time-dependent precession axis 
$\boldsymbol{\Omega}(t)$.

For identical envelopes $\epsilon_R(t)=\epsilon_B(t)=:\epsilon(t)$ as in Eq.~\eqref{eq:addition}, the $y$-component of the precession axis vanishes and the excited state occupation can be determined analytically as
\begin{align}
\label{eq:nt}
n(t)=&\frac 12\bigg[1-\cos\bigg(\int\limits_{-\infty}^t dt'\,
2\epsilon(t') \cos(\Delta t')\bigg)\bigg] ~.
\end{align}
In the limit $t\to\infty$ the integral in Eq.~\eqref{eq:nt} 
becomes the Fourier transform $\mathcal{F}[f](\omega=0)$ 
of $f(t)$, evaluated at the two-level transition frequency. This proves analytically that -- irrespective of the driving strength, no excited state population exists after the pulse unless there is overlap between the dichromatic excitation spectrum and the fundamental transition of the quantum emitter.

To illustrate this, Figure~\ref{fig:1s}(a) presents the dynamics of an ideal 2LS under dichromatic excitation with Gaussian pulses.
This shows transient excited state population which, however, vanishes again towards the end of the pulse, i.e.~the overall accumulated pulse area does indeed cancel almost completely. 
The small but finite residual occupation can be explained by the nonzero overlap between the tails of the Gaussians and the fundamental transition.
Consequently, coherent Rabi-like oscillations with unity population inversion can still be obtained, albeit at much larger intensities, as depicted in Figure~\ref{fig:1s}(b). 
However, this obviously defeats the purpose of employing the dichromatic excitation scheme.

The transient excited state occupation is key to understanding how significant population inversion can still be obtained even if the exciting pulse has no spectral overlap with the transition of the emitter: 
Any additional interaction or dissipation can interfere with the complete cancellation of the pulse area, and thus lead to a finite population inversion after the pulse. 
For example, in a laser-driven QD, the interaction with phonons induces incoherent thermalization dynamics in the instantaneous laser-dressed state basis, unlocking excited state population up to  $\sim 50\%$.

In this Letter, we propose and experimentally demonstrate an alternative, externally controllable approach to dichromatic pulsed excitation (DPE). To obtain large stationary occupations of the excited state we employ asymmetric dichromatic excitation with red and blue-detuned pulses with different intensities. 
For $\epsilon_B(t)\neq\epsilon_R(t)$, the Bloch sphere precession axis $\boldsymbol{\Omega}(t)$ then has a finite time-dependent $y$-component, adding more degrees of freedom to the coherent dynamics.
For suitable parameter choices, such asymmetric DPE can result in Bloch sphere trajectories that coherently evolve towards the excited state at long times.

We experimentally verify our insights by characterizing the dynamics and quality of the scattered photons under DPE of a solid-state 2LS.
As discussed in detail in the following, our results confirm a maximal population transfer fidelity of approximately 50\% under symmetric DPE, owing to incoherent phonon-induced dynamics. Further, we show that coherent dynamics with a population inversion of 80\% are achievable through an asymmetric weighting of the red and blue components of the dichromatic pulse. We conclude our study by analyzing the quality of the resulting photons in terms of the degree of multi-photon suppression and Hong-Ou-Mandel (HOM) visibility.

\begin{figure}
\includegraphics[width=.5\textwidth]{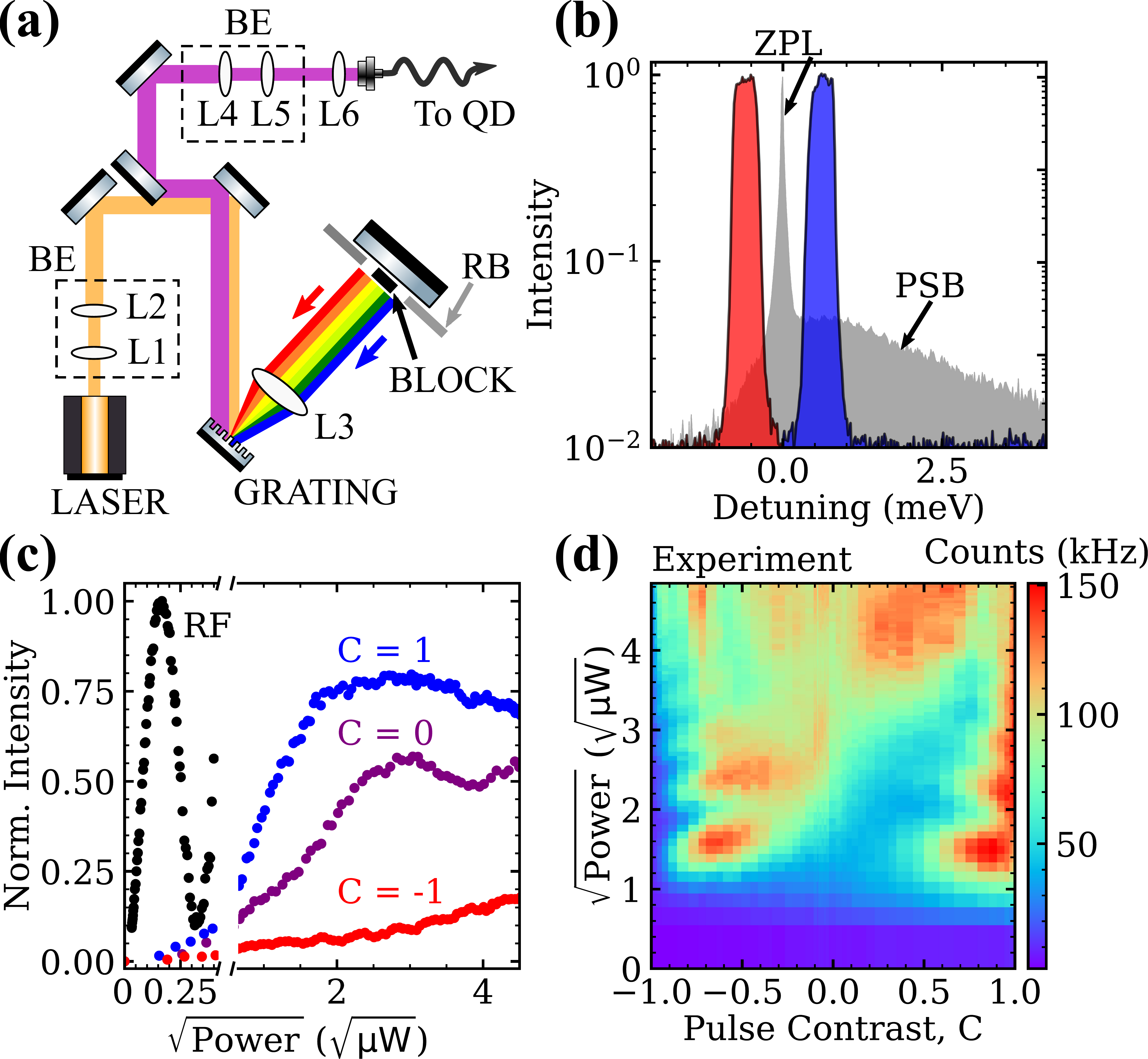}
\caption{\textbf{Experimental setup, spectral properties, and QD emission intensity under dichromatic pulse excitation.}
\textbf{(a)} Schematic of the folded $4f$ experimental setup for pulsed excitation, with
L: Lenses, BE: Beam expander, RB: Razor blades.
\textbf{(b)} Spectra of the excitation laser (red and blue) and the QD absorption (mirrored from the emission spectra, grey), as measured on the spectrometer. 
Spectral filtering with a bandwidth of $120\,\mu\mathrm{eV}$ after the collection fibre isolates the zero-phonon line (ZPL) from the phonon sideband (PSB) and excitation laser.
\textbf{(c)} Emission count rate as a function of excitation power using a single resonant pulse (RF), the red sideband ($C=-1$), blue sideband ($C=1$) and an equally weighted combination of the two ($C=0$), normalized to the maximum intensity obtained via RF excitation.
\textbf{(d)} Experimental data on the emission count rate as a function of pulse area and pulse contrast, defined as the weighted difference between the integrated intensity of the red and blue sidebands. 
Changes in the pulse contrast are implemented by varying the width of blue or red sidebands.
}
\label{fig:2s}
\end{figure}

As a solid-state 2LS for the dichromatic excitation experiments, we use the negatively-charged exciton transition, $X^{1-}$, of a charge-tunable, planar cavity InGaAs QD sample~\cite{malein_screening_2016,koong_fundamental_2019}. 
Figure~\ref{fig:2s}(a) shows the experimental setup to generate the dichromatic pulses for excitation.
A mode-locked laser with $80.3\,\mathrm{MHz}$ repetition rate and $160\,\mathrm{fs}$ pulse width is sent to a folded $4f$ setup, which consists of lenses, beam expanders (BE), a grating, a set of two motorized razor blades (RB) and a beam block.
The RB control the overall spectral width of the diffracted beam while a beam block placed between them removes the undesired frequency component resonant with the zero-phonon line, simultaneously ensuring phase-locking. 
After back reflection on a mirror, the remaining light recombines on the same grating and gets coupled into an optical fibre before exciting the QD.
Figure~\ref{fig:2s}(b) depicts an example of the spectra of the excitation laser and the absorption profile of the QD, detuned from the zero-phonon-line (ZPL) at $\omega_0=1.280\,\mathrm{eV}\,(968.8\,\mathrm{nm})$, measured using a spectrometer with $\sim 30\,\mu\mathrm{eV}$ resolution.
The spectrum of the QD shows an atomic-like zero-phonon line (ZPL), along with a broad, asymmetric phonon sideband (PSB) arising primarily from interaction with longitudinal acoustic (LA) phonons~\cite{koong_fundamental_2019,brash_light_2019}.
The excitation laser spectrum shows the spectral width and the separation of the red and blue sideband of $0.5\,\mathrm{meV}$ and $1.2\,\mathrm{meV}$, respectively.
We define the pulse contrast $C$ of the dichromatic pulse as a function of the integrated intensity of the red ($I_R$) and blue ($I_B$) sideband, as $C = (I_B-I_R)/(I_B+I_R)$. 
Finally, after filtering on the ZPL, the scattered photons are detected on a superconducting nanowire single photon detector with $\sim 90\%$ detection efficiency at $\sim 950\,\mathrm{nm}$.

We first compare the experiment results for symmetric dichromatic driving ($C=0$), blue-detuned excitation ($C=1$), and red-detuned excitation ($C=-1$) with that obtained via pulsed resonant fluorescence (RF).
These results are depicted in Figure~\ref{fig:2s}(c).
While we observe the expected Rabi oscillation under RF, we record much higher emission intensities at $C=1$ than at $C=-1$, consistent with findings from previous studies, and corresponding to phonon-assisted excitation~\cite{ardelt_dissipative_2014,quilter_phonon-assisted_2015}.
Contrary to the expected minimal state occupation for an ideal 2LS under symmetric dichromatic excitation at $C=0$ (c.f. Figure~\ref{fig:1s}), we observe a population inversion fidelity of $\approx 50\%$ at a pulse area of $\sim 20\,\pi$.
We attribute this to the unavoidable electron-phonon interaction: as discussed, phonon-induced thermalization allows occupations of $\lesssim 50\%$, compared to only vanishingly small levels for a dissipation-less 2LS.

\begin{figure}
    \includegraphics[width=0.5\textwidth]{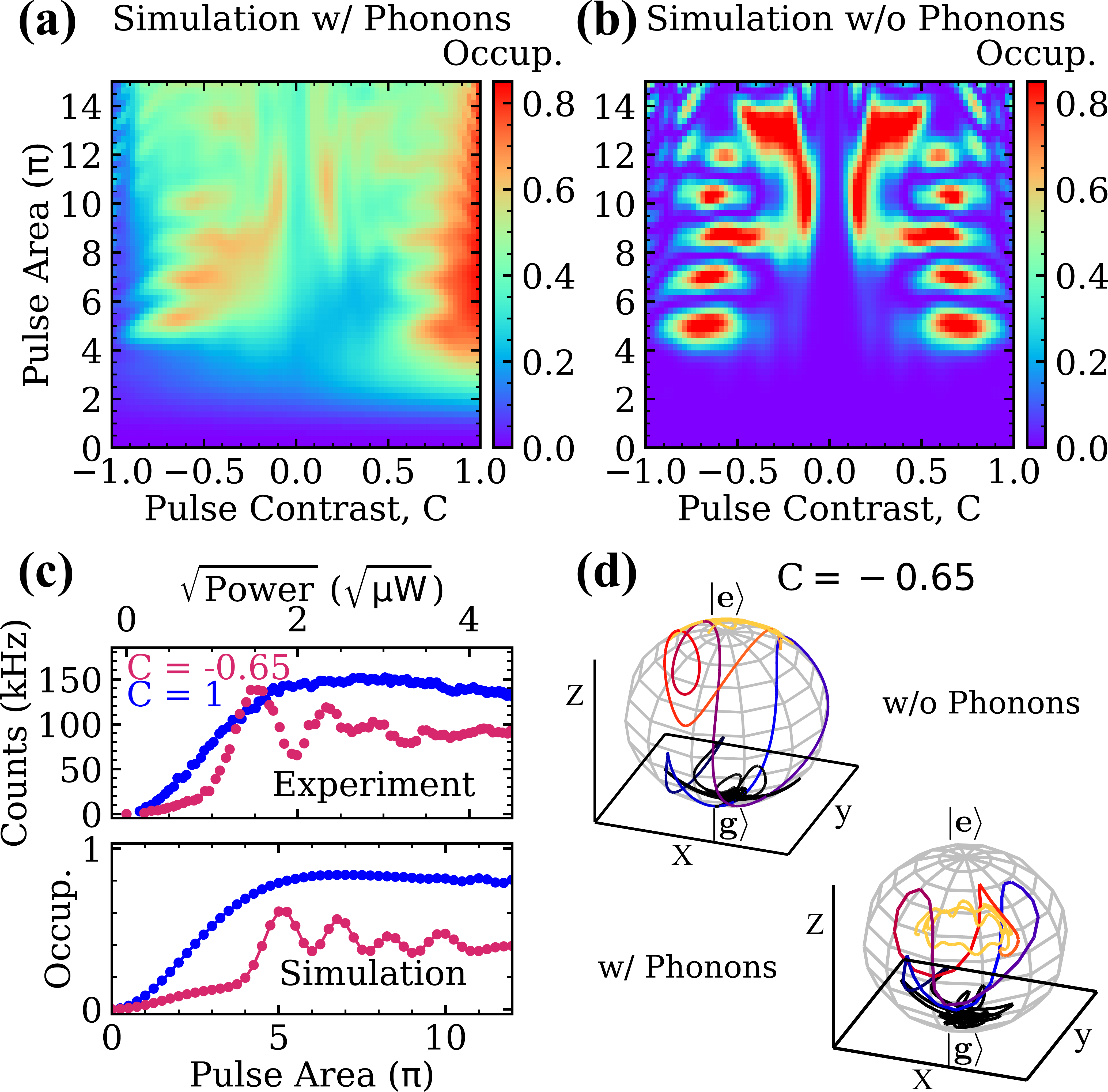}
    \caption{\textbf{Coherent dynamics of a 2LS under symmetric and asymmetric dichromatic excitation.}
    \textbf{(a, b)} Simulated (with phonons, a and without phonons, b) data on the emission count rate as a function of pulse area and pulse contrast, defined as the weighted difference between the integrated intensity of the red and blue sidebands. 
    \textbf{(c)} Comparison between the experimental data (top) and the simulated data in the absence of phonons (bottom) at $C=-0.65$ (magenta) and $C=1$ (blue).
    \textbf{(d)} Simulated dynamics of the first oscillation maxima (pulse area of $5\,\pi$), for asymmetric dichromatic excitation ($C=-0.65$) from calculations without (top) and with (bottom) phonon coupling, indicate complete (incomplete) population inversion without (with) phonon coupling. Darker colors represent earlier times. 
    The 2LS is initialized in the ground state $\Ket{g}$.
    }
    \label{fig:3s}
\end{figure}

We now proceed to characterize the dynamics of the system under asymmetric DPE.
To achieve this, an additional beam block mounted on a motorized translation stage is added in front of the RB to allow independent control of the width of the red or blue sideband.
The excitation pulse is split via a 99/1 fibre beam splitter, with the low power channel sent to the spectrometer to estimate the pulse contrast and the higher power channel used to excite the QD.
Figure~\ref{fig:2s}(d) shows the experimental measured emission count rate as a function of pulse contrast and excitation power.
We compare the experimental data with simulations using a numerically exact real-time path-integral formalism~\cite{cygorek_nonlinear_2017} with parameters typical of GaAs QDs~\cite{krummheuer_pure_2005} and employing a pair of rectangular driving pulses. 
We refer to Section I in the Supplementary Materials~\cite{[{See Supplemental Material at [URL will be inserted by publisher] for details on the simulation methods and parameters, quantum dot (QD) source, performance of the QD under monochromatic resonant fluorescence and phonon-assisted excitation, dynamics of the emission under dichromatic pulse excitation with various pulse parameters and multi-level systems, as well as a comparison of the two-photon visibility of the scattered photons between the monochromatic and dichromatic scheme, which includes Refs.~\cite{kuhlmann_charge_2013,scully_quantum_1997,markfox2006,iles-smith_limits_2017,morreau_phonon_2019,Scholl2019,muller_-demand_2014,tonndorf2015single,srivastava2015optically,he2015single,koperski2015single,branny2017deterministic,kumar2016resonant,errando-herranz_resonance_2020}}] SupMat} (SM) for full simulation parameters.  
The simulation, taking into account of the exciton-phonon coupling in Figure~\ref{fig:3s}(a) shows close qualitative agreement with the experimental data.
For comparison, the dynamics obtained in the absence of exciton-phonon coupling is depicted in Figure~\ref{fig:3s}(b).
Figures~\ref{fig:3s}(a) and (b) both feature oscillations in the excited state occupation at $C\approx \pm 0.65$, indicating their coherent nature. The non-oscillatory revival of the population inversion at $C=1$, as well as a higher inversion efficiency at $C\approx 0.8$ where the blue sideband dominates (compared to $C=-0.65$), are found only in the presence of phonon coupling. This confirms the presence of phonon-assisted excitation.
Figure~\ref{fig:3s}(c) shows the line-cuts of the experimental and simulated data from Figure~\ref{fig:2s}(d) and \ref{fig:3s}(a) taken at $C=-0.65$ and $1$, respectively; this indicates good qualitative agreement between experiment and theory.

To better illustrate the coherent dynamics of the asymmetric pulses, in Figure~\ref{fig:3s}(d) we present simulated 2LS population dynamics on the Bloch sphere for a pulse area up to the first coherent oscillation in Figure~\ref{fig:3s}(c) at $C=-0.65$.  
In the absence of dissipation (top), the state of the 2LS remains pure and is constrained to the surface of the Bloch sphere. The nontrivial spiralling trajectory is a consequence of the time-dependent $x$ and $y$ components of the effective electric field associated with the asymmetric pulse. In this particular instance, the trajectory evolves towards the excited state located at the north pole.
When the interaction with phonons is accounted for (bottom), the system features mixed-state dynamics that are no longer restricted to the surface of the Bloch sphere. Qualitatively, the spiralling trajectory still looks similar to that of the phonon-free case. However, now the excited state is no longer reached.
Rather, the projection onto the $z$-axis gives a final excited state occupation of $\approx 60\%$. Note that this value is lower than the measured 80\% inversion fidelity, likely due to a slight mismatch in the pulse shape between simulation and experiment.
In any case, our combined results indicate that for $C\approx -0.65$ phonons certainly quantitatively affect the dynamics but dominating coherent oscillations nonetheless survive. 
In contrast, for positive pulse contrast, the higher maxima of the coherent oscillations are strongly suppressed by the interaction with phonons. This qualitative difference in behaviour between positive and negative pulse contrast is attributable to the differing spectral overlap of the dichromatic pulse pair with the QD's phonon side band (cf.~Figure~\ref{fig:2s}).

Richer and even more complex dynamics emerges when moving beyond the case of a 2LS.  In Sections VII and VIII of the SM~\cite{SupMat} we present a range of spectroscopic results from more complex multi-level solid-state systems, however, a full exploration of those systems under DPE is beyond the scope of the present study.

\begin{figure}
\includegraphics[width=0.5\textwidth]{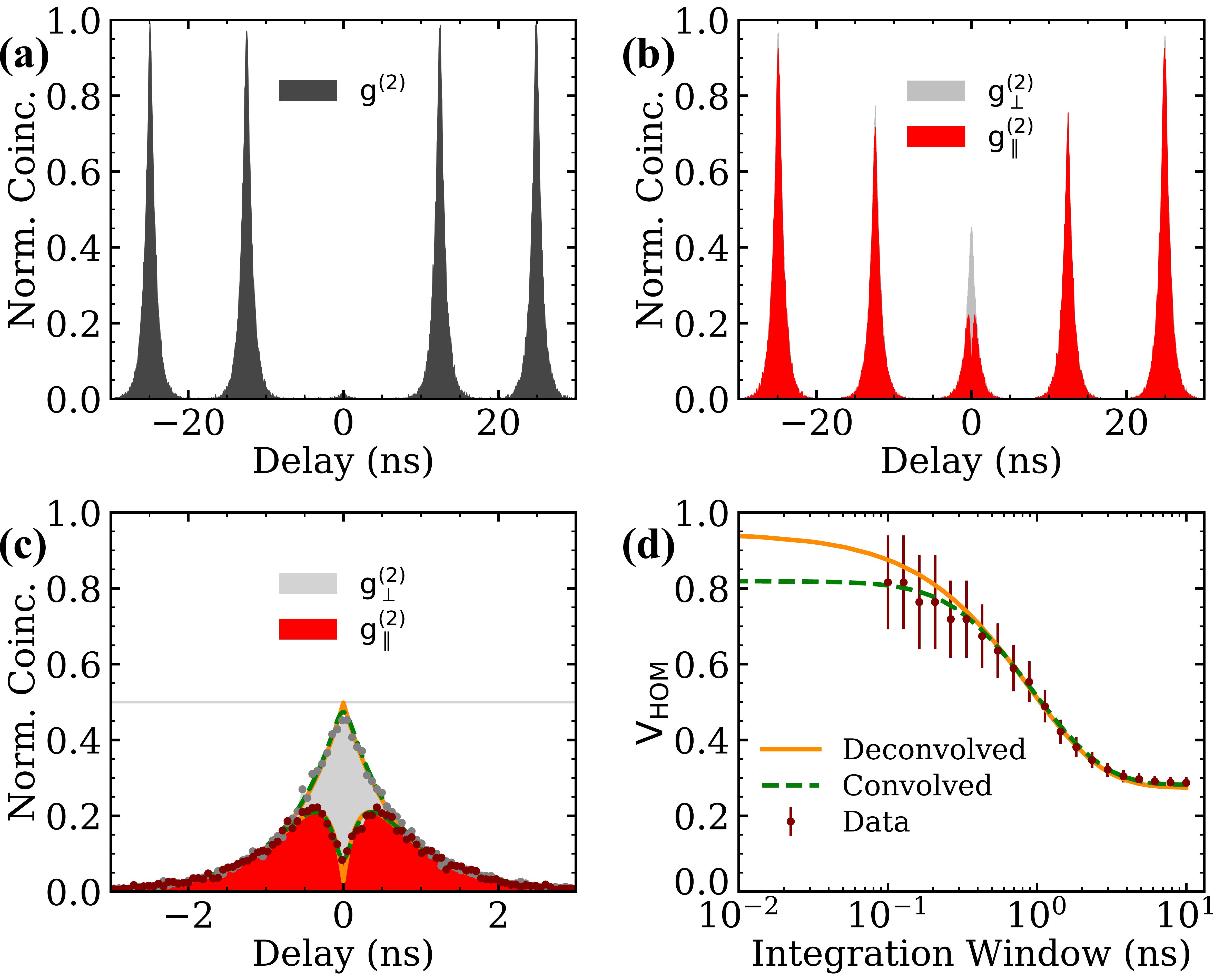}
\caption{\textbf{Single photon purity and indistinguishability under asymmetric dichromatic excitation at $C=-0.65$.}
\textbf{(a)} Measurement of the second order correlation function gives $g^{(2)}(0)=0.016\,(1)$.
\textbf{(b)} Two-photon interference of consecutively scattered photons delayed by $12.5\,\mathrm{ns}$, prepared in cross ($g^{(2)}_\perp$) and parallel ($g^{(2)}_\parallel$) polarizations.
\textbf{(c)} Close-up of the zero delay peak for $g^{(2)}_\parallel$ reveals a dip, due to temporal filtering from our detectors.
Dashed (green) and solid (orange) lines represent the convolved and the de-convolved fit to the experimental data (solid circles), respectively.
\textbf{(d)} Two-photon interference visibility $\rm V_{HOM}$ as a function of the integration time window for $g^{(2)}_{\parallel(\perp)}$ around $\tau = 0$.
The solid (dashed) line is obtained from integrating the convolved (de-convolved) fit in (c).
}
\label{fig:4}
\end{figure}

Having identified the pulse contrast and excitation power to optimize the emission count rate, we proceed to characterize the single photon performance from our QD under DPE.
By sending the photons into a Hanbury-Brown and Twiss interferometer, we observe multi-photon suppression of $g^{(2)}(0)=0.016\,(1)$, indicating near perfect single photon emission, as shown in Figure~\ref{fig:4}(a).
We then measure the indistinguishability of the scattered photons via HOM interference between two consecutively emitted photons at a time delay of 12.5~ns.
The figure of merit here is the two-photon interference visibility $\rm V_{HOM}$, determined by sending the photons into an unbalanced Mach-Zehnder interferometer with an interferometric delay of 12.5~ns to temporally match the arrival time of subsequently emitted photons on the beam splitter.
Figure~\ref{fig:4}(b) and (c) show the normalized HOM histogram as a function of time delay $\tau$ between detection events for photons prepared in cross ($g^{(2)}_\perp$) and parallel ($g^{(2)}_\parallel$) polarizations within a $60$~ns window and a $6$~ns wide zoom into the central peak, respectively. 
This close-up on the co-polarized $g^{(2)}_\parallel$ peak near the zero-delay illustrates the characteristic dip.
We fit the experimental data with the function $g^{(2)}_\parallel(\tau)=0.5\,\exp(-\tau/T_1)[1-\mathrm{V_{HOM}^{deconv.}}\times \exp(-\tau/\tau_C)]$~\cite{legero_time-resolved_2003,nazir_overcoming_2009,gold_two-photon_2014,reimer_overcoming_2016}, convolved with a Gaussian instrument response function (bandwidth of $0.168\,\mathrm{ns}$), where the independently measured lifetime is $T_1= 687\,(3)\,\mathrm{ps}$.
This yields a de-convolved visibility of $\mathrm{V_{HOM}^{deconv.}} = 0.95\,(1)$ and a $1/e$ width of $\tau_C=0.33\,(2)\,\mathrm{ns}$. 
The signature dip around the zero delay, usually present under non-resonant pumping and resonant two-photon excitation schemes, indicates deviation from the transform limit and thus imperfect photon wave packet coherence.
The width of the dip corresponds to the characteristic time of $T_2^*=2\tau_C=0.66\,(4)\,\mathrm{ns}$ for the inhomogeneous broadening of the emitter due to pure dephasing or timing jitter in emission ~\cite{legero_time-resolved_2003,patel_quantum_2010}.
We speculate that this may be dominated by phonon-induced dephasing, as we observe a narrower dip under phonon-assisted excitation while noting its absence under strict monochromatic resonant excitation. See Section III and IV in SM~\cite{SupMat} for the corresponding experimental evidence and discussion.
Figure~\ref{fig:4}(d) shows $\rm V_{HOM}$ as a function of integration window around $\tau = 0$ for temporal filtering of events between detection.
Temporal post selection~\cite{nazir_overcoming_2009} increases the raw visibility, $\rm V_{HOM}$ from $0.29\,(2)$ to $0.81\,(12)$ when narrowing the integration time window from 10~ns to 0.1~ns, respectively.
Integrating the fit function to $g^{(2)}_{\parallel(\perp)}$ (solid lines) gives a maximum convolved (de-convolved) visibility of  $\rm V_{HOM}=0.81\,(0.95) $.
The presence of residue coincidences around the zero delay in the histogram for scattered photons under DPE indicates the effect of finite time jitter and dephasing in the photon coherence, rendering the scheme partially coherent.

In summary, we have shown that, counter-intuitively, symmetric dichromatic excitation is unsuitable for achieving coherent population control of quantum emitters. 
Specifically, it suffers from excitation inefficacy due to cancellation of the accumulated pulse area, and the inversion efficiency scales with the spectral overlap of the driving pulses with the emitter resonance. 
This nullifies the purported advantage of separating the spectrum of the driving field from the emitter zero-phonon line for background-free photon extraction. Recognizing this problem, we demonstrate that a simple adjustment in the relative weighting of the red and blue-detuned pulses is sufficient to improve the population inversion efficiency whilst maintaining minimal spectral overlap. Unity population inversion is then possible for an ideal 2LS, and we have measured 80\% inversion efficiency with our QD sample. 
The presence of intensity oscillations under asymmetric driving demonstrate the coherent nature of the observed dynamics, yet those dynamics deviate from canonical Rabi oscillations and intrinsically feature non-trivial and complex Bloch-sphere trajectories. 
Our work has further experimentally demonstrated near perfect multi-photon suppression and high levels of photon indistinguishability (via temporal filtering) for such an asymmetric dichromatic excitation approach. This provides a new route to coherently excite quantum emitters, opening the prospect of background-free single photon extraction with suitably optimized cavity-coupled photonic solid-state devices~\cite{iles-smith_phonon_2017,gustin_pulsed_2018,denning_phonon_2020}.

\begin{acknowledgments}
This work was supported by the EPSRC (Grants No. EP/L015110/1, No. EP/M013472/1, and No. EP/P029892/1), the ERC (Grant No. 725920), and the EU Horizon 2020 research and innovation program under Grant Agreement No. 820423. 
B. D. G. thanks the Royal Society for a Wolfson Merit Award and the Royal Academy of Engineering for a Chair in Emerging Technology. 
E.M.G. acknowledges financial support from the Royal Society of Edinburgh and the Scottish Government. 
Y. M. acknowledge the support from Chongqing Research Program of Basic Research and Frontier Technology (No.cstc2016jcyjA0301).
The authors in KIST acknowledge the support by IITP grant funded by the Korea government (MSIT) (No. 20190004340011001)
\end{acknowledgments}

\bibliography{reference}

\end{document}


\title{Supplementary Materials for "Coherent Dynamics in Quantum Emitters under Dichromatic Excitation"}

\author{Z. X. Koong}
\email[Correspondence: ]{zk49@hw.ac.uk}
\affiliation{
 SUPA, Institute of Photonics and Quantum Sciences, Heriot-Watt University, EH14 4AS, United Kingdom
}
\author{E. Scerri}
\email[Current address: Leiden University, P.O. Box 9504, 2300 RA Leiden, The Netherlands]{}
\affiliation{
 SUPA, Institute of Photonics and Quantum Sciences, Heriot-Watt University, EH14 4AS, United Kingdom
}
\author{M. Rambach}
\email[Current address: Centre for Engineered Quantum Systems, School of Mathematics and Physics, University of Queensland, QLD 4072 Australia]{}
\affiliation{
 SUPA, Institute of Photonics and Quantum Sciences, Heriot-Watt University, EH14 4AS, United Kingdom
}
\author{M. Cygorek}
\author{M. Brotons-Gisbert}
\affiliation{
 SUPA, Institute of Photonics and Quantum Sciences, Heriot-Watt University, EH14 4AS, United Kingdom
}
\author{R. Picard}
\affiliation{
 SUPA, Institute of Photonics and Quantum Sciences, Heriot-Watt University, EH14 4AS, United Kingdom
}
\author{Y. Ma}
\affiliation{
College of Optoelectronic Engineering, Chongqing University of Posts and Telecommunications, Chongqing 400065, China
}
\author{S. I. Park}
\affiliation{
 Center for Opto-Electronic Materials and Devices Research, Korea Institute of Science and Technology, Seoul 02792, Republic of Korea
}
\author{J. D. Song}
\affiliation{
 Center for Opto-Electronic Materials and Devices Research, Korea Institute of Science and Technology, Seoul 02792, Republic of Korea
}
\author{E. M. Gauger}
\author{B. D. Gerardot}
\email[Correspondence: ]{b.d.gerardot@hw.ac.uk}
\affiliation{
 SUPA, Institute of Photonics and Quantum Sciences, Heriot-Watt University, EH14 4AS, United Kingdom
}
\email{zk49@hw.ac.uk}
\begin{abstract}
This document provides the details on the simulation methods and parameters, quantum dot (QD) source, performance of the QD under monochromatic resonant fluorescence and phonon-assisted excitation, dynamics of the emission under dichromatic pulse excitation with various pulse parameters and multi-level systems, as well as a comparison of the two-photon visibility of the scattered photons between the monochromatic and dichromatic scheme.
\end{abstract}
\date{\today}
\maketitle

\section{Simulation method and parameters}
For the simulations presented in the main text we model the quantum dot
including the coupling $H_\textrm{phonon}$ to longitudinal acoustic phonons 
by the total Hamiltonian 
\begin{subequations}
\begin{align}
\label{eq:tot}
H_\textrm{tot}=&H+H_\textrm{phonon} \\
H=&\frac \hbar 2\big( f^*(t) |e\rangle\langle g| 
+ f(t)|g\rangle\langle e|\big)\\
H_\textrm{phonon}=& 
\sum_\mathbf{q}\hbar\omega_\mathbf{q} b^\dagger_\mathbf{q} b_\mathbf{q}
+\sum_\mathbf{q} \hbar \gamma_\mathbf{q}
\big(b^\dagger_\mathbf{q}+b_\mathbf{q}\big)|e\rangle\langle e|,
\end{align}
\end{subequations}
where $|g\rangle$ and $|e\rangle$ are the ground and excited states of the quantum dot (QD),
respectively, $b^\dagger_\mathbf{q}$ is the creation operator of a phonon
in mode $\mathbf{q}$, $\hbar\omega_\mathbf{q}$ is the energy of mode 
$\mathbf{q}$, and $\gamma_\mathbf{q}$ describes the strength of the coupling
between phonon mode $\mathbf{q}$ and the exiced QD state.

Using a real-time path integral method~\cite{PhysRevB.96.201201}, the dynamics induced by the total Hamiltonian $H_\textrm{tot}$
is solved numerically exactly, i.e., without any approximation other than a
finite time discretization. 
The influence of the phonons, which are assumed to be initially in thermal 
equilibrium at temperature $T=4$ K, is uniquely determined by the
phonon spectral density $J(\omega)$. For deformation potential coupling to
longitudinal acoustic phonons
\begin{align}
J(\omega)&=\sum_\mathbf{q} \gamma_\mathbf{q}^2 \delta(\omega-\omega_\mathbf{q})\\
&=\frac{\omega^3}{4\pi^2\rho\hbar c_s^5}
\bigg(D_e e^{-\omega^2a_e^2/(4c_s^2)} - D_h e^{-\omega^2a_h^2/(4c_s^2)}\bigg)^2.
\end{align}
We use standard parameters~\cite{PhysRevB.71.235329} for a GaAs-based
QD with electron radius $a_e=3.0$ nm, hole radius $a_h=a_e/1.15$, 
speed of sound $c_s=5110$ m/s, density $\rho=5370$ kg/m$^3$ and electron 
and hole deformation potential constants $D_e=7.0$ eV and $D_h=-3.5$ eV, 
respectively.

To approximate the pulses used in the experiment, we assume a rectangular 
shape in the frequency domain. 
The red and blue-detuned pulses each has a full-width-at-half-maximum (FWHM) of $\Gamma=0.4$ meV and is detuned by $\Delta=0.6$ meV from the transition energy of the two-level system.
As in the experiment, different overall 
intensities for red and blue-detuned pulses are 
implemented via different spectral 
widths of the rectangles $W_R$ and $W_B$ while the heights of the rectangles
are chosen to be the same. In the time domain these pulses take the form
\begin{align}
f(t)=&\epsilon_R(t)e^{-i\Delta t}+\epsilon_B(r)e^{i\Delta t} \\
\epsilon_{R/B}(t)=&\frac{A}\pi \frac{\sin [W_{R/B}(t-t_0)/(2\hbar)]}{t-t_0},
\end{align}
where $t_0$ is the time corresponding to the center of the pulse and $A$ is
the pulse area for a single resonant pulse in absence of QD-phonon interactions.

\section{Quantum dot Source Characterization}

\begin{figure*}
\includegraphics[width=1\textwidth]{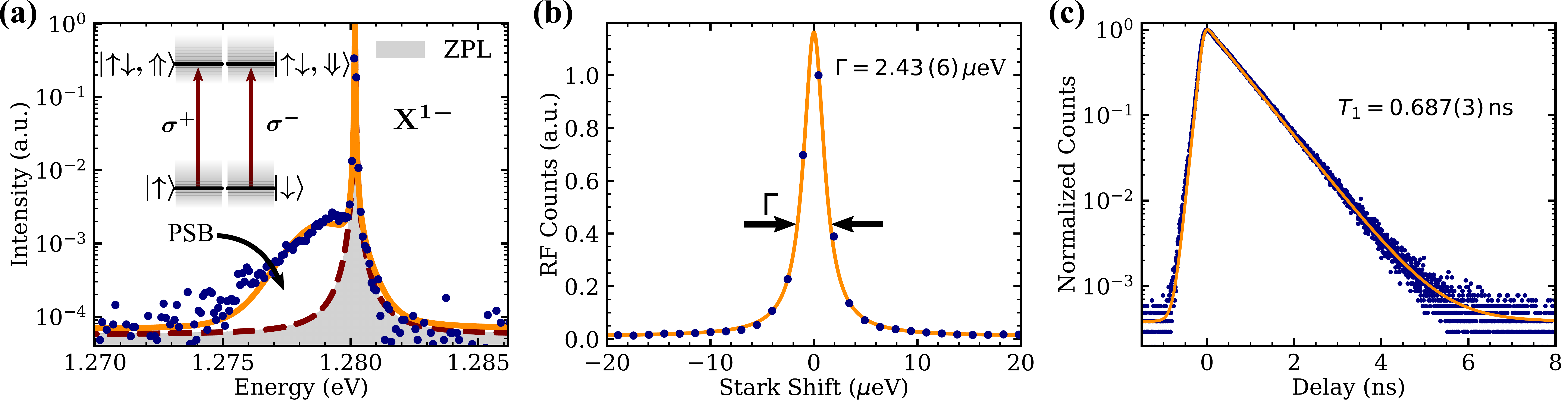}
\caption{\textbf{QD source characterization.}
\textbf{(a)} Emission spectrum of the QD under CW resonant excitation (resonance fluorescence, RF) showing the zero-phonon line (ZPL, dashed) and the phonon sideband (PSB). 
Inset shows the energy level schematic for a negatively-charged exciton, $\rm X^{1-}$ under zero external magnetic field.
The solid line is produced using the phonon and material parameters and model from Ref.~\cite{koong_fundamental_2019}.
\textbf{(b)} Detuning spectrum of the $\rm X^{1-}$ transition at low excitation power.
The Lorentzian fit (solid line) gives a full-width-at-half-maximum (FWHM, $\Gamma$) of $\Gamma=2.43\,(6)\,\mu\mathrm{eV}$.
\textbf{(c)} Time-resolved emission from the $\rm X^{1-}$ transition under pulsed RF excitation. 
The convolved (with the instrument response function with FWHM of 160 ps) fit (solid line) shows a single exponential decay, giving an excited state lifetime of $T_1=0.687\,(3)\,\mathrm{ns}$.
}
\label{fig:1s}
\end{figure*}

Figure~\ref{fig:1s}(a) shows the emission spectrum of the QD under resonant continuous wave (CW) excitation, showing a narrow zero-phonon line (ZPL, shaded) and a broad phonon-sideband (PSB), originated from relaxation from the phonon-dressed states.
The fit (orange solid line) is obtained from the polaron model using previously cited parameters~\cite{koong_fundamental_2019}, which gives a ZPL fraction of $\approx 92\%$.
The schematic of the energy level of the $\rm X^{1-}$ transition in zero magnetic field (inset of Figure~\ref{fig:1s}A) shows two degenerate ground spin states ($\ket{\uparrow}$ and $\ket{\downarrow}$), each coupled to their corresponding excited state, i.e. $\ket{\uparrow}\Leftrightarrow\ket{\uparrow\downarrow,\Uparrow}$ and $\ket{\downarrow}\Leftrightarrow\ket{\uparrow\downarrow,\Downarrow}$.
Here, the single (double) arrows refer to electron (heavy-hole) spin state. 
Each transition has a well-defined optical selection rule such that it can be optically coupled with right ($\sigma^+$) or left ($\sigma^-$) circular polarized light.
Keeping the frequency of the excitation laser fixed at $\omega_0=1.280\,\mathrm{eV}$, we scan through the resonance of the QD via d.c. Stark tuning to measure the linewidth of the scattered QD photons.
A Lorentzian fit to the detuning spectra of the QD in Figure~\ref{fig:1s}(b) under weak excitation gives a full-width-at-half-maximum (FWHM) of $\Gamma=2.43\,(6)\,\mu\mathrm{eV}$.
Figure~\ref{fig:1s}(c) shows the time-resolved lifetime measurement of the emission under pulsed monochromatic resonant excitation at $\pi$-pulse. 
A single-sided exponential decay fit to the data (convolved with the instrument response function with FWHM of $160\,\mathrm{ps}$) reveals an excited state lifetime of $T_1=0.687\,(3)\,\mathrm{ns}$.
This corresponds to a transform-limited linewidth of $\sim 1\,\mu\mathrm{eV}$.
The deviation of measured linewidth $\Gamma$ from the transform-limited linewidth indicates the existence of pure-dephasing from the solid-state environment, possibly originating from charge and spin noise of the QD device~\cite{kuhlmann_charge_2013,kuhlmann_transform-limited_2015,malein_screening_2016}. 

\section{Pulsed monochromatic resonance fluorescence (RF)}
\label{sec:rf}
\begin{figure*}
\includegraphics[width=1\textwidth]{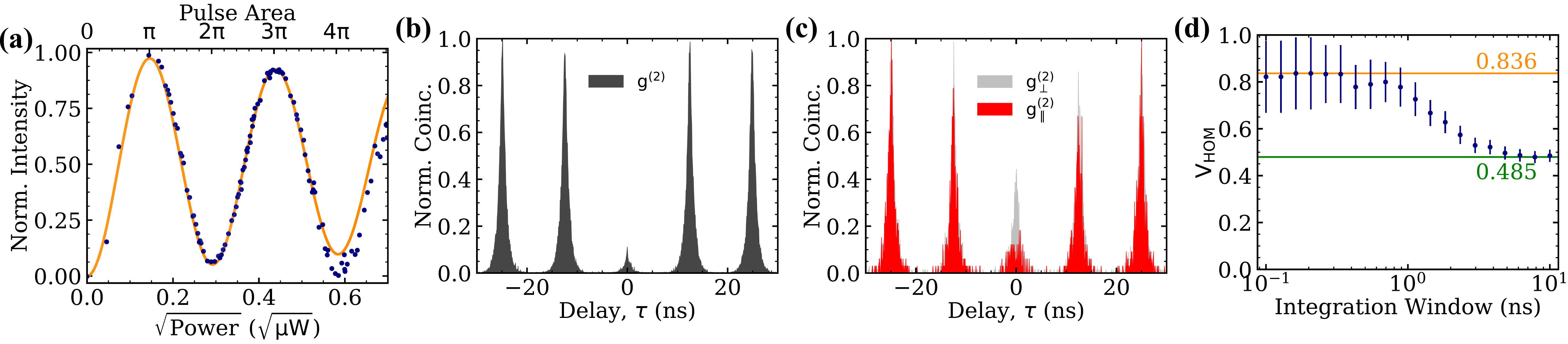}
\caption{\textbf{Monochromatic resonance fluorescence (RF).}
\textbf{(a)} 
Emission intensity as a function of square root of the excitation power of a single 14-ps Gaussian pulse, resonant to the ZPL at $\omega_0\,=\,1.280\,\mathrm{eV}$, normalized to the intensity at the $\pi$-pulse.
The fit (solid line) to the experimental data (circles) is derived from the pure dephasing model~\cite{scully_quantum_1997,markfox2006}.
\textbf{(b)} Intensity-correlation histogram of the scattered photons at $\pi$-pulse shows a multi-photon suppression of $g^{(2)}(0)=0.080\,(2)$.
\textbf{(c)} Two-photon interference histogram of the consecutively emitted QD photons at $\pi$-pulse, prepared in parallel ($g^{(2)}_\parallel$) and perpendicular ($g^{(2)}_\perp$) polarization measured using an unbalanced Mach-Zehnder interferometer.
\textbf{(d)} The visibility of the scattered photons as a function of integration window shows a post-selected visibility $\rm V_{HOM}$ of $0.84\,(15)$ and $0.49\,(3)$ at $100\,\mathrm{ps}$ and $10\,\mathrm{ns}$ integration windows, respectively.
}
\label{fig:2}
\end{figure*}

To benchmark the performance of the dichromatic pulse excitation (DPE) scheme, we perform pulsed monochromatic resonance excitation (resonance fluorescence, RF) on the same transition (and the same QD).
We optically excite the QD using a $\approx 14\,\mathrm{ps}$-width pulse (spectral bandwidth of $\approx 80\,\mu\mathrm{eV}$), and filter out the QD signal via polarization and spectral filtering to suppress the excitation laser spectrum.
Figure~\ref{fig:2}(a) shows the normalized intensity of the emission as a function of the square root of the excitation power.
We fit the data using the time-dependent excited state population function, derived from the pure dephasing model~\cite{scully_quantum_1997,markfox2006}, showing coherent Rabi oscillation as a function of pulse area.
Fixing the excitation power to a $\pi$-pulse, we perform intensity correlation and Hong-Ou Mandel (HOM)-type two-photon interference measurements on the scattered photons.
Due to the imperfect excitation laser rejection (signal-to-background of $\sim 20$), we obtain a multi-photon suppression of $g^{(2)}(0)=0.080\,(2)$ for the scattered photons, as shown in Figure~\ref{fig:2}(b).
In Figure~\ref{fig:2}(c, d), we observe a post-selected HOM visibility $\rm V_{HOM}$ of $0.84\,(15)$ and $0.49\,(3)$ at $100\,\mathrm{ps}$ and $10\,\mathrm{ns}$ integration windows, respectively.
The HOM visibility is computed as the ratio of two-photon interference of consecutive photons, prepared in parallel, $ g^{(2)}_\parallel$ and in perpendicular polarization, $ g^{(2)}_\perp$, which follows as
\begin{equation}
\mathrm{V_{HOM}} =1-g^{(2)}_\parallel/g^{(2)}_\perp.
\label{eqn:hom}
\end{equation}
With monochromatic resonant excitation, despite the higher $g^{(2)}(0)$ due to the imperfect suppression of the excitation laser, we observe a higher two-photon interference visibility ($\rm V_{HOM}=0.58\,(3)$), in comparison to the value obtained under DPE ($\rm V_{HOM}=0.29\,(2)$, see Figure~4 in main text) and under phonon-assisted excitation ($\rm V_{HOM}=0.19\,(1)$, see Figure~\ref{fig:3} in Section~\ref{sec:pa}).
This implies, while the DPE scheme benefits from the fact that polarization filtering is not needed for background-free single photon collection, the RF excitation technique is still preferred as a means to generate single photons with higher indistinguishability.

\section{Phonon-assisted excitation}\label{sec:pa}
\begin{figure*}
\includegraphics[width=1\textwidth]{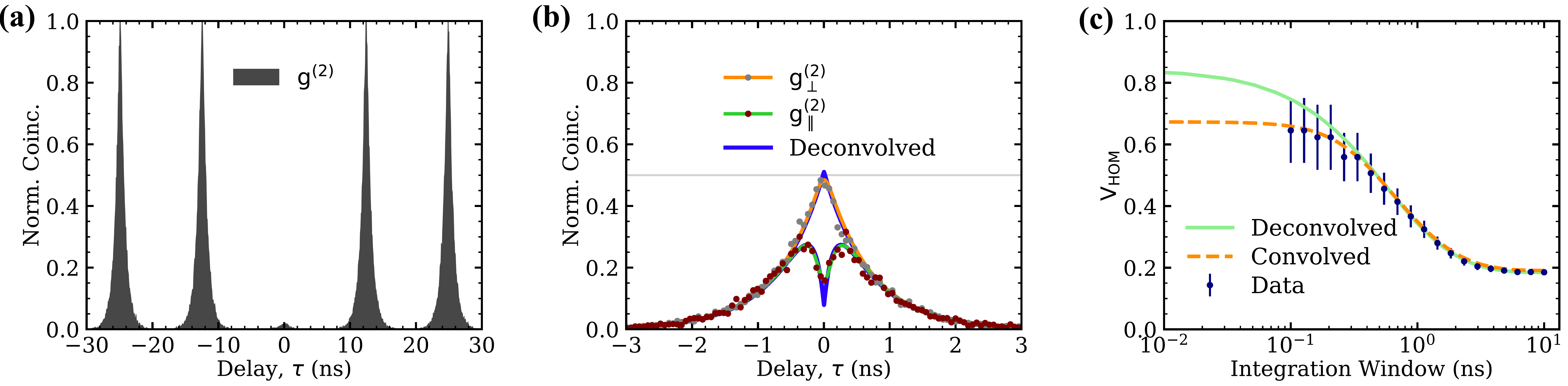}
\caption{\textbf{Phonon-assisted excitation on the same QD and transition.}
\textbf{(a)} Intensity-correlation histogram at pulse area of $\approx 10\pi$ shows a multi-photon suppression of $g^{(2)}(0)=0.025\,(1)$.
\textbf{(b)} Close-up of the zero delay peak in the two-photon interference of consecutively scattered photons delayed by $12.5\,\mathrm{ns}$, prepared in cross ($g^{(2)}_\perp$) and parallel ($g^{(2)}_\parallel$) polarizations.
The blue solid lines represent the deconvolved fit to the experimental data (solid circles), respectively. 
\textbf{(c)} Two-photon interference visibility $\rm V_{HOM}$ as a function of the integration time window for $g^{(2)}_{\parallel(\perp)}$ around $\tau = 0$.
The solid (dashed) line is obtained from integrating the convolved (deconvolved) fit in (b).
}
\label{fig:3}
\end{figure*}

Figure~\ref{fig:3} demonstrates the performance of the QD (for the same transition) under phonon-assisted excitation. 
The excitation laser pulse has a pulse width of 7~ps and is detuned $\approx 0.8\,\mathrm{meV}$ from the ZPL.
The excitation pulse area is $\approx 20\,\pi$, corresponding to saturation count rate.
The scattered photons are then spectrally filtered with the same $120\,\mu\mathrm{eV}$-bandwidth grating filter to suppress the scattered laser.
Despite large multi-photon suppression, giving $g^{(2)}(0)=0.025\,(1)$, due to the emission timing jitter that arises from the absorption of phonons assisting the population of the excited state, we observe a HOM visibility is slightly lower than that for the DPE and RF schemes, giving a post-selected HOM visibility of $\mathrm{V_{HOM}}=0.64\,(14)$ and $\mathrm{V_{HOM}}=0.19\,(1)$ at $100\,\mathrm{ps}$ and $10\,\mathrm{ns}$ integration windows, respectively.
In addition, we observe a narrower (and shallower) dip (giving a $1/e$ width of 158~ps) around the zero-delay in $g^{(2)}_\parallel$, as indicated in Figure~\ref{fig:3}(b), compared to that observed under DPE scheme ($1/e$ width of 330~ps). 
Fitting the dip with 
$g^{(2)}_\parallel(\tau) =0.5\exp{(-\tau/T_1)}[1-\mathrm{V_{HOM}^{deconv.}}\times\exp{(-\tau/\tau_C)}]$, where $T_1$ is the emitter lifetime and $\tau_C$ is the $1/e$ width of the zero-delay dip,
gives a maximum, deconvolved HOM visibility of $\mathrm{V_{HOM}^{deconv.}}=0.83\,(1)$, suggesting that dephasing due to the phonon-bath operates at a time scale way shorter than the bandwidth of our detection instrument response function (FWHM$= 160\,\mathrm{ps}$), as predicted in Ref.~\cite{iles-smith_limits_2017}.

\section{Two-photon Interference Visibility: Comparison between RF and DPE schemes}
\begin{figure*}
\includegraphics[width=1\textwidth]{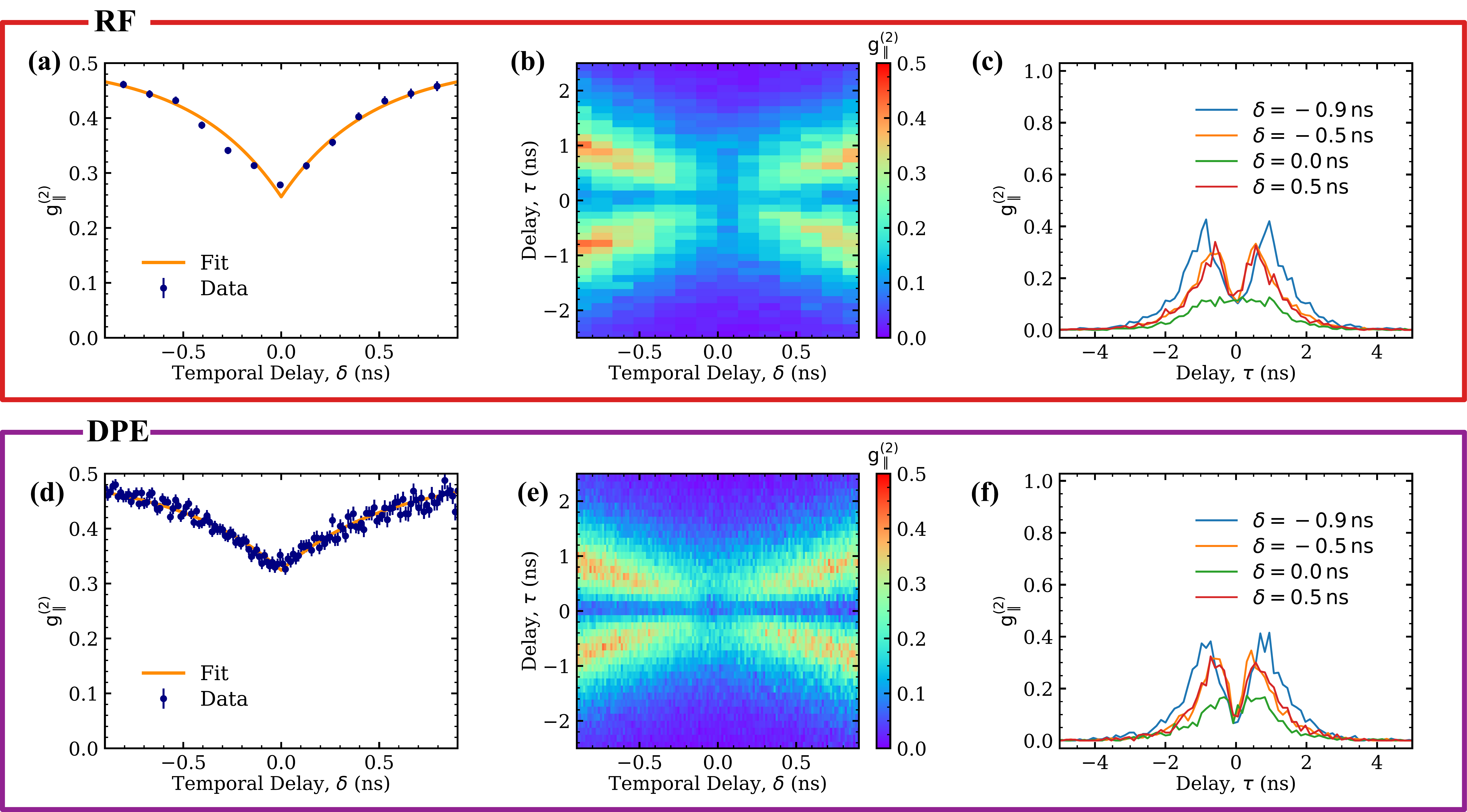}
\caption{\textbf{Comparison between the Hong-Ou-Mandel (HOM) interference under monochromatic resonant excitation (RF, a-c) and dichromatic excitation (DPE, d-f).} 
\textbf{(a,d)} HOM-type interference of two photons from consecutively scattered photons delayed by $12.5\,\mathrm{ns}$ given by $g^{(2)}_\parallel(0)$, integrated over a detection window of 10 ns, as a function of temporal delay $\delta$ between the input photons.
\textbf{(b,e)}  $g^{(2)}_\parallel(\tau,\delta)$ as a function of $\delta$ and detection time delay $\tau$, within a detection window of $6\,\mathrm{ns}$.
\textbf{(c,f)} Cross-section of the $g^{(2)}_\parallel(\tau,\delta)$ at a temporal delay of $\delta=-0.9$, -0.5, 0 and 0.5~ns.
}
\label{fig:4}
\end{figure*}

In this section, we report on the results of on the HOM visibility as a function of temporal delay between the arrival time of the two input photons on the beam splitter, $\delta$.
We render both input paths of the beam splitter indistinguishable in polarization ($\rm g^{(2)}_\parallel$), and measure the detection time delay $\tau$ between "click" events on the photon detectors for each $\delta$. 
We perform this measurement on the same transition and QD under both RF and DPE schemes.
Figure~\ref{fig:4} shows the comparison of the HOM visibilities between the RF (a-c) and DPE (d-f) schemes.
Figure~\ref{fig:4}(a) and (d) shows the normalized coincidence around the zero delay, $\rm g^{(2)}_\parallel(0)$, integrated over a window of 10 ns as a function of $\delta$.
When the two input photons perfectly overlap with each other on the beam splitter ($\delta=0$), we observe a minimum in $ g^{(2)}_\parallel(0)$. The data is fitted with a simple exponential function (${ g^{(2)}_\parallel}(\tau=0,\delta)=0.5\times\left(1-\mathcal{V}\exp(-|\delta |/T_2)\right)$) to extract the coherence time $T_2$ and the visibility of the HOM dip, $\mathcal{V}$, of the scattered photons.
We obtain a $T_2=0.457\,(27)\,\mathrm{ns}$ ($T_2=0.548\,(13)\,\mathrm{ns}$) and $\mathcal{V}=0.49\,(2)$ ($\mathcal{V}=0.35\,(1)$) for scattered photons under RF (DPE) excitation. 
The lower HOM visibility $\mathcal{V}$ for the DPE scheme, despite much higher signal-to-background ratio, is due to presence of the dip in the detection time histogram. This is evident in Figure~\ref{fig:4}(b, c, e, f).
Figure~\ref{fig:4}(b) and (e) show the 2D plot of the normalized coincidence as a function of both $\delta$ and $\tau$.
We observe a similar pattern reported in Ref.~\cite{legero_time-resolved_2003,patel_quantum_2010}, in which the presence non-vanishing dip around the zero detection time delay $\tau=0$ is due to either the timing jitter or pure dephasing mechanism.
Figure~\ref{fig:4}(c) and (f) show the coincidence histogram of ${ g_\parallel^{(2)}}(\tau,\delta)$ for $\delta=$-0.9, -0.5, 0 and 0.5~ns.
The appearance of the dip even at perfect overlap ($\delta=0$) in the DPE case with negligible background, is a signature of pure dephasing/timing jitter in the emission, which originates from phonon-induced dephasing~\cite{iles-smith_limits_2017,morreau_phonon_2019}.
In Section~\ref{sec:pa}, we observe a similar signature (narrow dip around the zero time delay $\tau$) in ${ g^{(2)}_\parallel}(\tau,\delta=0)$, which further confirms our claim that the HOM visibility suffers from the same phonon-induced dephasing mechanism in the DPE scheme. 
For an emission that is dephasing and jitter free, we expect the disappearance of the dip around $\tau=0$~\cite{kambs_limitations_2018}.
We attribute the disappearance of the dip for the RF case as a signature of jitter- or dephasing-free performance, and the non-vanishing coincidence ${ g^{(2)}_\parallel}(\tau=0,\delta=0)$ is solely due to the imperfect filtering of the background laser scattering in the collection.
With proper filtering to improve the signal-to-background ratio (ideally $>100$), we should be able to minimize these coincidences, giving close to unity indistinguishability~\cite{Scholl2019}.

\section{Dichromatic Pulses with different pulse parameters}
\begin{figure}
\includegraphics[width=0.5\textwidth]{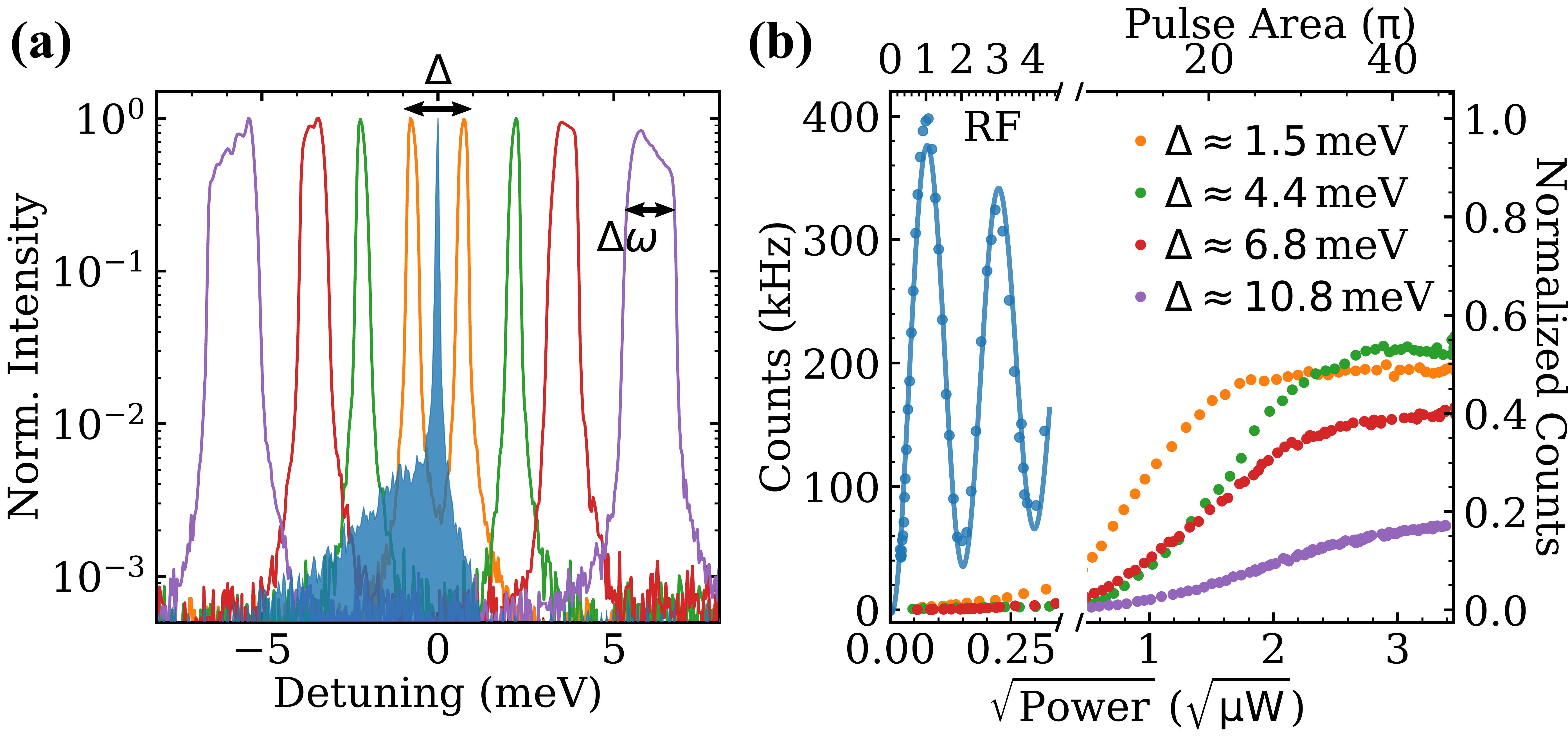}
\caption{
\textbf{Dichromatic excitation dependency on pulse parameters.}
\textbf{(a)} Spectra of the excitation laser, detuned from ZPL of the QD emission (shaded, blue) at 1.2730~eV for various pulse detuning, $\Delta$ and pulse width, $\Delta \omega$, with corresponding pairs in the same colour.
The spectral width and the amplitude of the red and blue-detuned pulses are kept the same.
The emission profile from the QD serves as a reference to highlight the spectral overlap between the excitation pulses with the phonon sideband at different pulse parameters.
\textbf{(b)} Emission count rate as a function of square root of the excitation power at dichromatic pulse detuning of $\Delta=0$ (RF), 1.5, 4.4, 6.8 and 10.8~meV.
The corresponding pulse widths are $\Delta \omega=0.05$ (RF), $0.24$, $0.24$, $0.65$ and $1.30\,\mathrm{meV}$, respectively.
The pulse area of $\pi$ corresponds to square root of the excitation power needed to achieve maximum intensity under pulsed RF.
}
\label{fig:dpe_pulse_parameters}
\end{figure}

This section explores the population inversion efficiency of a solid-state two-level system for different pulse parameters under DPE.
Here, we address the negatively-charged exciton ($\rm X^{1-}$) transition of a different QD.
The two pulse parameters: pulse width, $\Delta \omega$, and pulse detuning, $\Delta$, are used to characterize the pulse shapes. 
They are defined as the spectral width of the red/blue-detuned pulse and the detuning between red and blue-detuned pulses, respectively.
To reduce experimental complexity, we vary the pulse width and detuning symmetrically, keeping the red and blue-detuned components of the dichromatic pulses the same throughout.

Figure~\ref{fig:dpe_pulse_parameters} shows the emission spectra and the detected count rates from pulsed RF and DPE at various pulse parameters.
Here, we vary the thickness of the beam block and the separation of the razor blades in the pulse strecher (see Figure.~2(a) in the main text) to remove the particular spectral components in the original 160~fs (corresponds to spectral bandwidth of $\sim 11\,\mathrm{meV}$) laser pulse.
The excitation laser spectra used for dichromatic excitation is illustrated in Figure~\ref{fig:dpe_pulse_parameters}(a), along with the resonantly driven emission spectrum from the $\rm X^{1-}$ transition under pulsed RF at $\pi$-pulse, in order to highlight the spectral overlap between the laser pulses and the broad phonon-sideband.
Figure~\ref{fig:dpe_pulse_parameters}(b) shows the emission count rate as a function of square root of the average excitation power for various DPE pulse parameters.
The pulse area is normalized to the optical power needed for a $\pi$-pulse under pulsed monochromatic RF (pulse width of $\sim 35\,\mathrm{ps}$ and bandwidth of $\Delta \omega=0.05\,\mathrm{meV}$).
Under pulsed RF, we observe Rabi oscillation in emission intensity as excitation power increases beyond the $\pi$-pulse.
The fit (solid blue line) to the experimental data (RF, blue circles) is derived from the same model used for the fit in Figure~\ref{fig:2}(a).
Unlike the monochromatic RF case, we observe a sigmoid-like saturation curve in the emission intensity.
We observe reduction of saturation intensities, below $0.5$ times the intensity at $\pi$-pulse under RF, and an increase in the excitation power needed to reach saturation as the pulse detuning $\Delta$ increases beyond 1.5~meV.
The observed reduction in the saturation intensity with $\Delta$ is consistent with the reduction in population inversion efficiency under monochromatic phonon-assisted excitation at large detuning~\cite{ardelt_dissipative_2014}, confirming the impact of phonon-mediated preparation and excitation-induced dephasing~\cite{ramsay_phonon-induced_2010,morreau_phonon_2019}.
The anomaly at $\Delta=4.4\,\mathrm{meV}$ can be attributed to either dominant phonon-assisted driving than the dichromatic driving or experimental imperfection in the excitation.
For instance, any imbalanced in components of the red and blue-detuned pulses, slight detuning of the dichromatic pulses from the ZPL of the QD emission and chirping in femtosecond pulses introduced by the dispersion of the fibre would deviate from the theoretical behaviour.
Nevertheless, the experimental evidence shows that the dynamics of the emission is sensitive to the excitation dichromatic pulse details (frequency detuning, pulse contrasts, pulse widths and pulse detuning). 
Hence, extra care has to be taken when selecting pulse parameters for the dichromatic pulses, ideally avoiding femtosecond pulses (with pulse detuning $\Delta \gtrapprox 2\,\mathrm{meV}$) if the excitation pulses are made to propagate in optical fibres to minimize any possible pulse chirping effects.

\section{Dichromatic excitation on the neutral exciton }
\begin{figure*}
\includegraphics[width=1\textwidth]{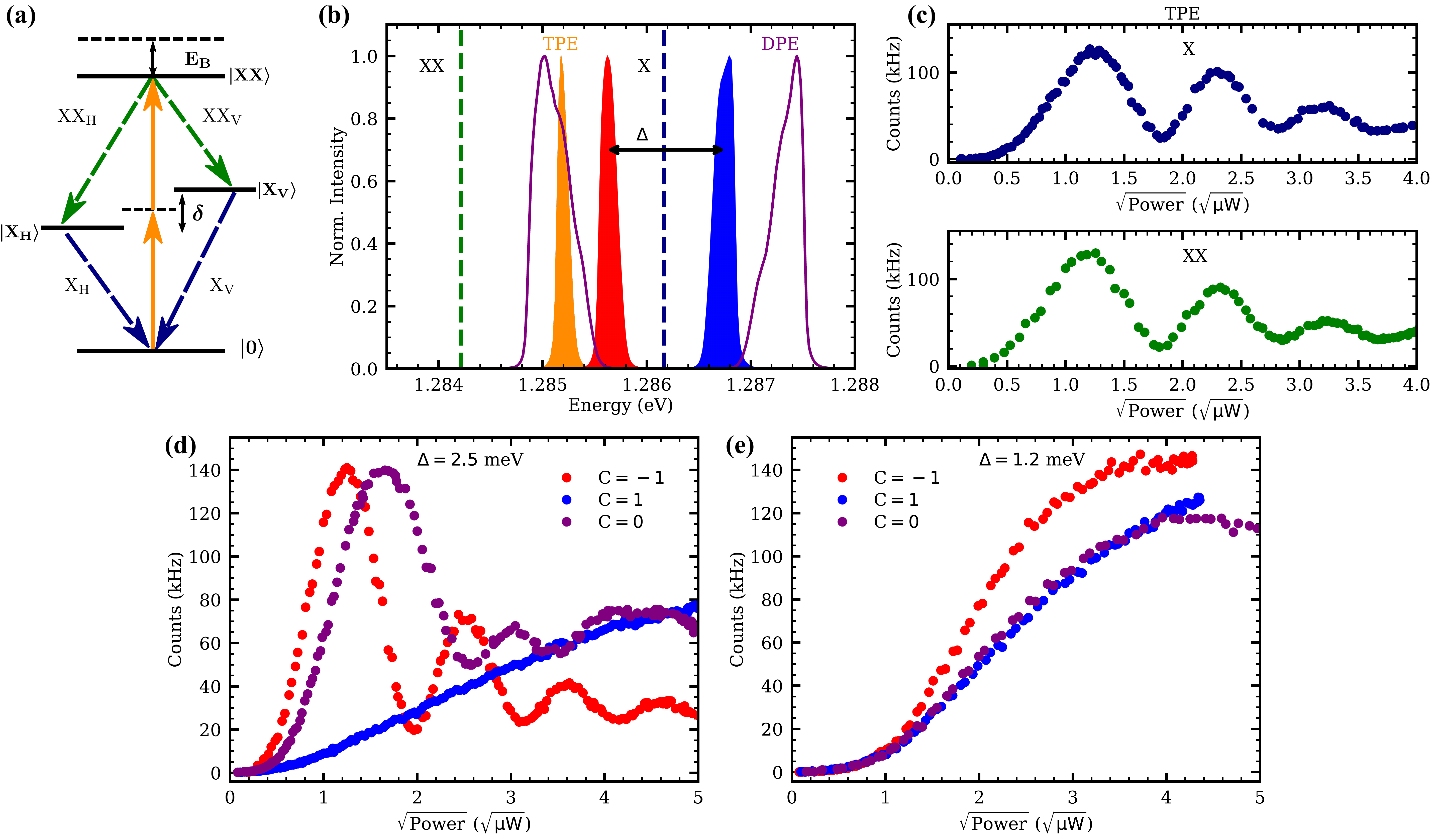}
\caption{\textbf{Dichromatic excitation on the neutral exciton $\rm X$ transition.}
\textbf{(a)} Energy level schematic of the biexciton-exciton (XX-X) cascaded system.
 $\delta$: exciton fine structure splitting. $E_B$: biexciton binding energy.
\textbf{(b)} Spectrum of the excitation laser for resonant two-photon excitation (TPE, orange) and the dichromatic pulse excitation (DPE, purple). 
The red and blue-detuned (from the X transition) sideband of the dichromatic pulse are separated by a pulse detuning, $\Delta$.
The biexciton-exciton state (XX) and the exciton-ground state transitions are indicated as green and blue dashed lines, respectively.
\textbf{(c)} The detected count rate of the scattered photons from the X (top, blue) and the XX (bottom, green) transitions  as a function of excitation power under TPE.
\textbf{(d, e)} The detected count rate of the scattered photons from the X transition as a function of excitation power under red-detuned ($C=-1$, red), blue-detuned ($C=1$, blue) and dichromatic ($C=0$, purple) excitation for pulse detuning of $\Delta=2.5$ (d) and $1.2~\mathrm{meV}$, respectively.
Rabi oscillation is observed when the red-detuned sideband overlaps with the two-photon resonance (orange curve in (b)).
}
\label{fig:6}
\end{figure*}

In this section, we perform DPE on the neutral exciton $\rm X$ transition.
Here, we consider balanced dichromatic pulses, with (dichromatic) pulse contrast $C=(I_B-I_R)/(I_B+I_R)\approx 0$, where $I_R$ and $I_B$ are the integrated intensity of the red and blue-detuned (from the resonance of the X transition) component of the dichromatic pulses, respectively.
Figure~\ref{fig:6}(a) shows energy level schematic of the four-level biexciton-exciton (XX-X) cascade system.
Upon excitation into the biexciton state $\ket{\rm XX}$, a cascaded radiative decay from biexciton state $\ket{\rm XX}$ to the vacuum ground state $\ket{0}$ is initiated via either of the intermediate neutral exciton states $\ket{\rm X_{H (V)}}$.
This generates a pair of polarization-entangled, orthogonally polarized photon pairs consisting of emission from both the biexciton-exciton (XX) and the exciton-vacuum (X) states transitions, distinguished via polarization (in the horizontal (H) or vertical (V) linear polarized basis) and difference in emission energy equal to the biexciton binding energy, $E_B$.
Figure~\ref{fig:6}(b) illustrates the laser spectrum for both the DPE (at pulse detuning of $\Delta=1.2$ and $2.5\,\mathrm{meV}$) and the resonant two-photon excitation (TPE).
The two-photon resonance lies at the half the energy difference between the X and XX transition, as indicated by the dashed lines, which gives a biexciton binding energy of $E_B=1.95\,(1)\,\mathrm{meV}$. 
The exciton fine structure splitting, independently measured via time resolved lifetime measurement, gives $\delta=19.6\,(1)\,\mu\mathrm{eV}$.
We resolve one of the exciton fine structures $\ket{\rm X_{H (V)}}$ by adjusting the linear polarizer in the collection to the polarization axis of the desired transition, while keeping maximal suppression in the excitation laser by calibrating the orientation of the linear polarizer in the excitation accordingly.
We spectrally filter either transitions before detecting the photons on a SNSPD. 

Figure~\ref{fig:6}(c) shows the emissions from the two transitions, observed simultaneously under TPE, as a function of the excitation power.
As demonstrated in previous literature~\cite{muller_-demand_2014} , we observe Rabi oscillation in both X and XX emissions, which enables coherent manipulation of the state occupation of the excitonic states.
Surprisingly, when employing dichromatic driving on the same transitions with red-detuned pulses overlapping with the two-photon resonance  ($\Delta=2.5\,\mathrm{meV}$), we observe similar Rabi oscillation, shown as purple circles in Figure~\ref{fig:6}(d).
Here, we speculate that unlike the solid-state two-level system (negatively-charged exciton, $\rm X^{1-}$), the contribution from two-photon resonance driving (red-detuned) to the state population inversion dominates over the phonon-assisted driving (blue-detuned).
We validate this by showing that the Rabi oscillation observed at $C=0$, is similar to that observed under TPE when we drive the X transition solely with the red-detuned pulses ($C=-1$).
Additionally, we observe lower emission intensities when excitation laser only consists of the blue-detuned component  ($C=1$) of the dichromatic pulses.
These evidences confirm our hypothesis, indicating a deviation from the expected outcome from the solid-state two-level system (c.f. Figure~2 and 3 in the main text) when dealing with multi-level system.

As we decrease the dichromatic pulse detuning to $\Delta=1.2\,\mathrm{meV}$ such that there is minimal overlap between red-detuned pulses with the two-photon resonance, we observe the disappearance of the Rabi oscillation when both red and blue-detuned component of the dichromatic pulses are present.
In addition, we observe a higher emission intensity when it is driven with the red-detuned pulses, compared to the lower detected count rate under phonon-assisted driving using blue-detuned pulses.
These results are illustrated in Figure~\ref{fig:6}(e).
It is interesting to note that for the blue-detuned driving ($C=1$), while still having lower the emission intensity as the red-detuned driving ($C=-1$), it shows sign of saturating at excitation power beyond $25\,\mu\mathrm{W}$.
This indicates that even when there is no overlap between the excitation pulses and the two-photon resonance, for a biexciton-exciton cascade system, the contribution from two-photon resonant driving ($C=-1$) dominants over the phonon-assisted driving ($C=1$) in affecting the state dynamics.
This adds further complexity in exploiting the DPE technique to coherently drive of the neutral exciton, $\rm X$ transition.
Further modeling would be beneficial to understand the physics behind this phenomena and to potentially utilize it as a tool for coherent single photon generation for multi-level atom-like system.

\section{Dichromatic Excitation of Single-Photon Emitters in Two-Dimensional Material}
\begin{figure*}
\centering
\includegraphics[width=0.95\textwidth]{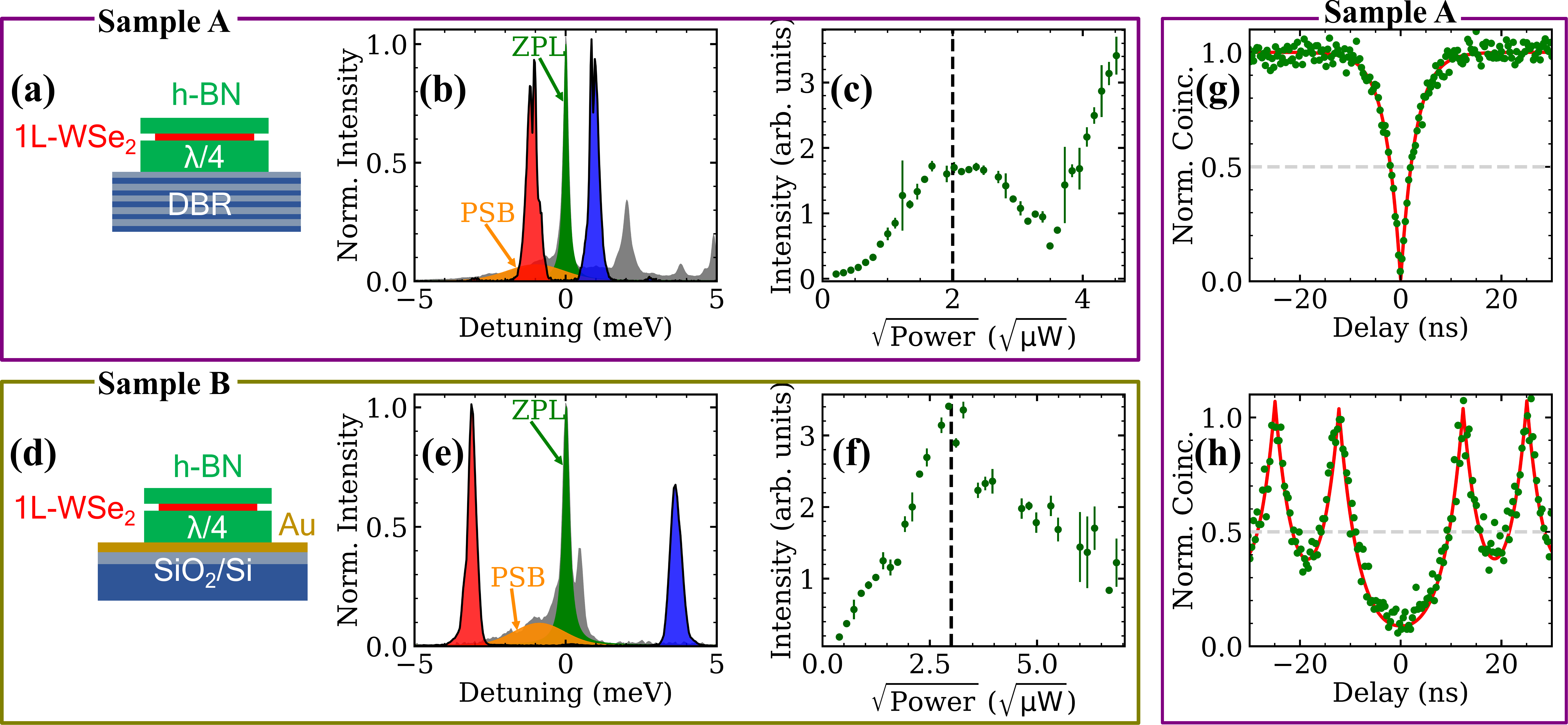}
\caption{
\textbf{Dichromatic excitation on single-photon emitters hosted by monolayer tungsten diselenide (1L-WSe$_{2}$) in Sample A (a-c) and B (d-e).}
\textbf{(a, d)} Sample structure, highlighting the hBN-1L-WSe$_{2}$-hBN heterostructure on top of a planar cavity, formed by either the distributed Bragg reflectors (DBR) (a) or gold (Au) mirror on SiO$_{2}$/Si substrate (d). 
\textbf{(b, e)} Photoluminescence emission profile (grey) of the single-photon emitters under non-resonant continuous wave excitation at 532~nm (2.33~eV), showing the zero-phonon-line (ZPL, green) and the phonon-sideband (PSB, orange).
The red and blue shaded regions indicate the red and blue-detuned (from the ZPL) components of the dichromatic laser, respectively.
The other peaks correspond to ZPL of other single-photon emitters within the collection spot.
\textbf{(c, f)} Emission intensity of the ZPL as a function of excitation power of the dichromatic laser.
The black dashed line corresponds to the $\pi$-pulse.
\textbf{(g, h)} Normalized coincidence from Hanbury-Brown and Twiss measurement of the ZPL in Sample A show suppressed multi-photon emission probability of $g^{(2)}(0)= 0.001 $ and $g^{(2)}(0)= 0.01$ under non-resonant (750~nm) continuous wave (g) and pulsed excitation (h), respectively.
}
\label{fig:2d}
\end{figure*}

In this section, we study the implementation of DPE on single photon emitters (SPEs) in two-dimensional materials.
In particular, we focus on SPEs hosted by monolayer tungsten diselenide (1L-WSe$_{2}$)~\cite{tonndorf2015single, srivastava2015optically, he2015single, koperski2015single,branny2017deterministic}, embedded in two different sample structures.

The architecture for the two samples, labeled as sample A and B, are illustrated in Figure~\ref{fig:2d}(a) and (d), respectively.
While both of them have the same heterostructure, which consists of 1L-WSe$_{2}$ encapsulated by few layers of hexagonal boron nitride (h-BN), their sample structures differs in the planar cavity design.
For sample A, the heterostructure is placed on top of a 140~nm stopband flat distributed Bragg reflector (DBR) centred at $\approx 710$~nm ($ 1.7463$~eV) with a 6~nm thick bottom h-BN flake acting as a spacer, forming a $\lambda/4$ planar cavity at $\lambda = 780$~nm ($1.5895$~eV). 
In contrast, for sample B, the heterostructure is placed on top of a gold mirror with a bottom hBN flake of 59.3~nm, creating a $\lambda/4$ planar cavity at $\lambda = 780$~nm ($ 1.58954$~eV). 
The photoluminescence emission from SPEs in both samples (grey, shaded), excited using the same non-resonant continuous wave source at 532~nm (2.33~eV), are shown in Figure~\ref{fig:2d}(b) and (e).
Their emission profile are detuned from the zero-phonon line (ZPL, green, shaded) at $1.6025$ and $1.5946$~eV, respectively.
Upon a close inspection of the emission spectra in Figure~\ref{fig:2d}(b), we observe emission peaks, which correspond to the ZPL from multiple emitters. 
The two peaks (ZPL and the peak beside it) in Figure~\ref{fig:2d}(e) belongs to the exciton fine structures of the same transition.
The dichromatic laser spectra are displayed alongside the SPEs emission spectra, with the red and blue-detuned (from the ZPL) laser components given by the red and blue shaded region, respectively.
The broad phonon-sideband (PSB, orange, shaded), detuned $\approx -0.9\,\mathrm{meV}$ from the ZPL, is indicated in both figures.
By computing the integrated intensities of the ZPL and PSB, we obtain a ZPL fraction of $\approx 65\,\%$ for both samples, typical for SPE in these materials at cryogenic temperature.
Subsequently, we filter out the ZPL using a grating-based spectral filter (FWHM$=0.296\,(1)\,\mathrm{meV}$) to suppress the laser sideband before it is detected on a spectrometer.
The emission intensity of the ZPL, as a function of excitation power for both samples, are shown in Figure~\ref{fig:2d}(c) and (f), respectively.
While the pulse parameters for the two dichromatic excitation differ (e.g. the dichromatic pulse detuning, $\Delta$ for the excitation on sample A and B are $\Delta=2.0$ and $6.7$~meV, respectively), we observe some form of oscillations.
Figure~\ref{fig:2d}(g) and (h) demonstrate suppressed multi-photon emission probability, $g^{(2)}(0)\sim 0$, from the spectrally filtered ZPL signal in Sample A, measured using a fibre-based Hanbury-Brown and Twiss interferometer, under continuous wave and pulsed non-resonant 750~nm excitation, respectively.
These results confirm the nature of single photon emission from these emitters.

While an accurate interpretation of the experimental data is currently unavailable due to the lack of clear quantum optical picture for these emitters, these results demonstrate coherent population driving of SPEs in 1L-WSe$_{2}$ under DPE, as an alternative to monochromatic resonant excitation~\cite{kumar2016resonant,errando-herranz_resonance_2020}.

\bibliography{reference}